\documentclass[10pt,twocolumn,letterpaper]{article}

\usepackage{wrapfig}
\usepackage{makecell}
\usepackage{tabularx}
\usepackage{booktabs}
\usepackage{multirow}

\newcommand{\gm}{GraphiMind}
\newcommand{\tci}{Textual Conversational Interface}
\newcommand{\gmi}{Graphical Manipulation Interface}

\newcommand{\cmark}[1]{{#1}}

\usepackage[pagenumbers]{cvpr} 

\usepackage[dvipsnames]{xcolor}

\definecolor{cvprblue}{rgb}{0.21,0.49,0.74}
\usepackage[pagebackref,breaklinks,colorlinks,citecolor=cvprblue]{hyperref}

\def\paperID{*****} 
\def\confName{CVPR}
\def\confYear{2024}

\title{GraphiMind: LLM-centric Interface for Information Graphics Design}

\author{Qirui Huang$^{1}$\quad Min Lu$^{1}$\quad Joel Lanir$^{2}$ \quad Dani Lischinski$^{3}$ \quad Daniel Cohen-Or$^{4}$ \quad Hui Huang$^{1}$\thanks{Corresponding author}\\
$^{1}$Shenzhen University \quad
$^{2}$University of Haifa \\
$^{3}$The Hebrew University of Jerusalem \quad
$^{4}$Tel Aviv University \\
{\tt\small $^{1,3,4}$\{qrhuang2021, lumin.vis,  danix3d, cohenor, hhzhiyan\}@gmail.com \quad $^{2}$ylanir@is.haifa.ac.il}
}

\begin{document}
\maketitle

\begin{abstract}

Information graphics are pivotal in effective information dissemination and storytelling. 
However, creating such graphics is extremely challenging for non-professionals, since the design process requires multifaceted skills and comprehensive knowledge.
Thus, despite the many available authoring tools, a significant gap remains in enabling non-experts to produce compelling information graphics seamlessly, especially from scratch.
Recent breakthroughs show that Large Language Models (LLMs), especially when tool-augmented, can autonomously engage with external tools, making them promising candidates for enabling innovative graphic design applications.
In this work, we propose a LLM-centric interface with the agent \textit{GraphiMind} for automatic generation, recommendation, and composition of information graphics design resources, based on user intent expressed through natural language.
Our GraphiMind integrates a \textit{Textual Conversational Interface}, powered by tool-augmented LLM, with a traditional \textit{Graphical Manipulation Interface}, streamlining the entire design process from raw resource curation to composition and refinement.
Extensive evaluations highlight our tool's proficiency in simplifying the design process, opening avenues for its use by non-professional users.
Moreover, we spotlight the potential of LLMs in reshaping the domain of information graphics design, offering a blend of automation, versatility, and user-centric interactivity.

\end{abstract}


\begin{figure*}
\centering
\includegraphics[width=\textwidth]{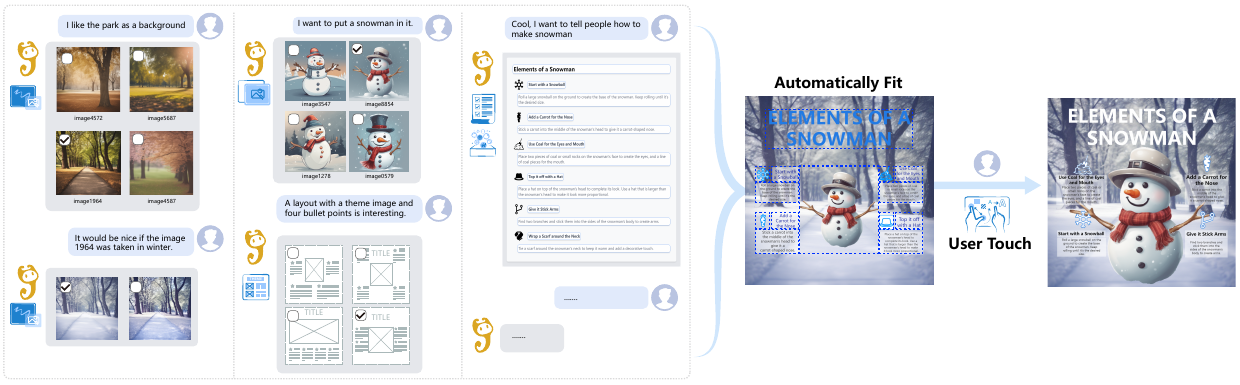}
\caption{\cmark{An Example of the Design Process in \textit{\gm}: users effortlessly communicate their design intention with the LLM agent in natural language, by which a wide range of core design assets are generated, including pivot figures, layouts, visual elements, and more. These resources can be seamlessly incorporated into a canvas, thereby facilitating the creation of information graphics with ease.}}
\label{fig:teaser}
\end{figure*}
\section{Introduction}\label{sec:intro}

Information graphics refers to visual representations that convey information in a clear and concise manner, by combining text and a variety of graphical elements in a visually appealing manner~\cite{smiciklas2012power}. Such graphics play an essential role in engaging storytelling and effective information communication~\cite{bateman2010useful, haroz2015isotype, harrison2015infographic}. Starting from an initial concept, designing an information graphic involves a multi-step process, from collection and preparation of information to detailed visual design and composition~\cite{Anjul2022_info}.  
The creation of information graphics demands creators to possess a comprehensive set of artistic and technical skills and knowledge spanning the entire process, making full-stack expertise essential. In recent years, advanced authoring tools have facilitated the design of information graphics~\cite{zhang2020dataquilt, coelho2020infomages, cui2021mixed}. However, most tools only focus on a part of the design process, assuming that the data content and desirable graphic elements are already available~\cite{chen2019towards, cui2019text, Anjul2022_info}. Thus, creating compelling information graphics, especially from scratch is still out of the reach of design novices. 
Moreover, these graphic design tools necessitate users to engage in direct manipulation of visual elements, thereby requiring the application of design principles and skills to achieve compelling results.

Large Language Models (LLMs) have been pivotal in advancing the field of Natural Language Processing (NLP)~\cite{devlin2018bert, brown2020language, chowdhery2022palm, touvron2023llama}, and now form the foundation of numerous popular products using text-based user interfaces, including the coding assistant Copilot\footnote{\url{https://github.com/features/copilot}}, and the popular ChatGPT chatbot\footnote{\url{https://openai.com/blog/chatgpt/}}. 
The rapid evolution of LLMs has facilitated significant strides in both text understanding and reasoning. These new capabilities pave the way for \textit{tool-augmented} LLM systems~\cite{parisi2022talm, mialon2023augmented, qin2023tool, ruan2023tptu}, where an LLM serves as a central controller to bridge between users and third-party tools, such as general computer vision and machine learning APIs~\cite{liang2023taskmatrix}, deep learning models~\cite{shen2023hugginggpt}, or domain specialized LLMs~\cite{wu2022promptchainer, hong2023metagpt}. Tool-augmented LLMs can automatically invoke relevant external tools based on the user's intent, achieved either through fine-tuning~\cite{schick2023toolformer} or by injecting tool-related information via in-context prompts~\cite{liang2023taskmatrix, gupta2023visual, hsieh2023tool}. These astounding advancements in LLMs motivate us to leverage their power in the domain of information graphics design. Unlike most of the non-design tasks current LLM systems work on (e.g., data analysis as highlighted by Zhang et al.~\cite{zhang2023data}), information graphics design presents a complex creative process involving multiple stages. 
It demands substantial human involvement and interaction throughout its iterative process, \cmark{which used to be achieved via different graphics manipulation software, such as Powerpoint, Adobe Illustrator, or other graphic authoring tools~\cite{Anjul2022_info}. Lately, several efforts have started to look at leveraging the power of LLMs in graphic design, e.g., Visual ChatGPT~\cite{wu2023visual}.} It's worth noting that text is the primary way humans express thoughts and ideas.
Furthermore, Large Language Models have demonstrated strong capabilities in text understanding. These two factors, when combined, introduce unique research problems, particularly when leveraging LLMs as text-based interfaces for information graphics design.

\begin{figure*}[htbp]
    \centering
    \includegraphics[width=0.98\textwidth]{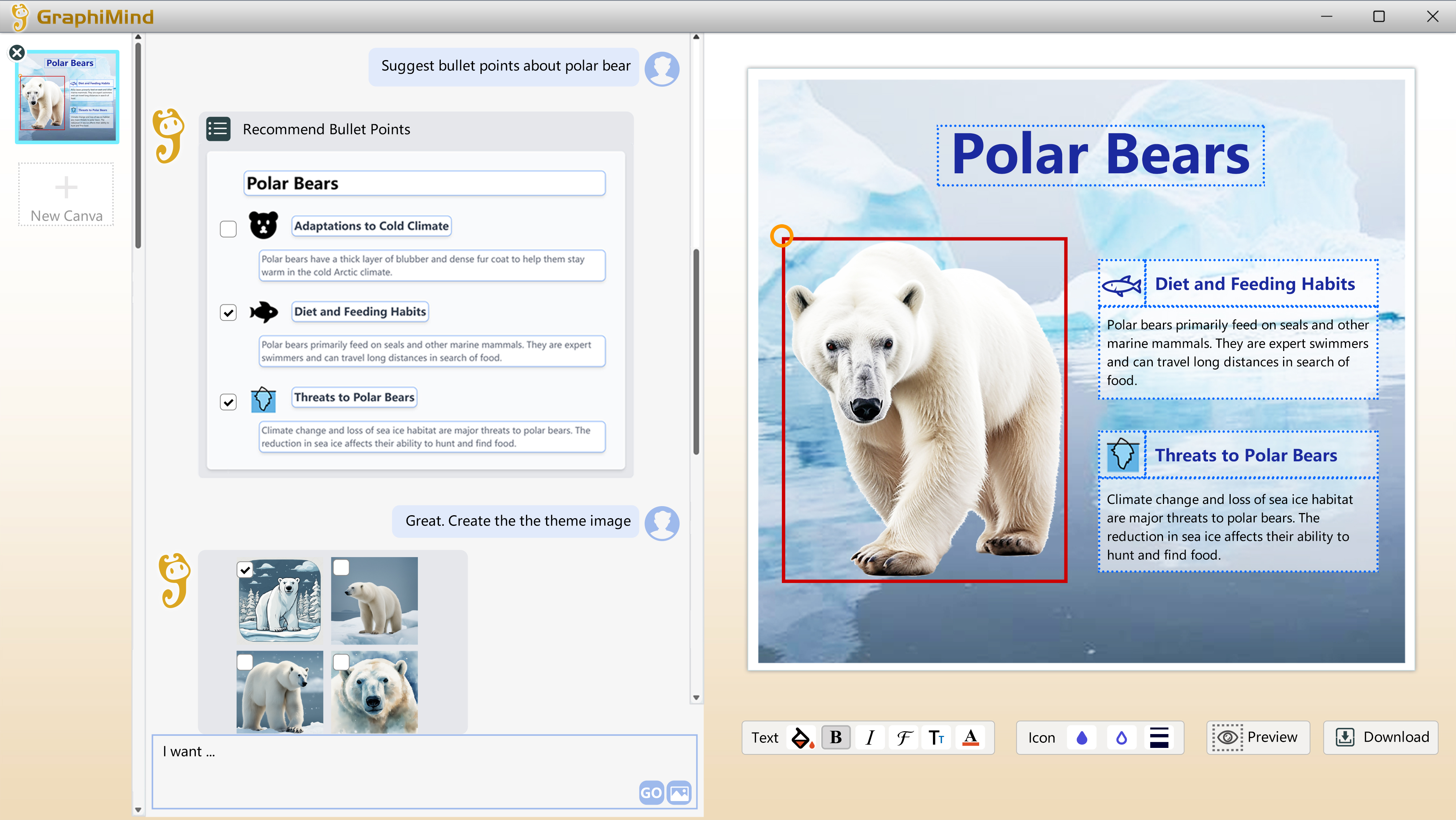}
    \caption{\cmark{The Interface of GraphiMind System: the system integrates a Textual Conversational Interface (on the left), enhanced by a tool-augmented LLM as an agent, with a Graphical Manipulation Interface (on the right). This integrated interface supports a range of essential design tasks in information graphics design from information collection to final adjustments.}}
    \label{fig:The UX interface.}
\end{figure*}

In this work, we present a novel LLM-centric user interface \textit{GraphiMind}, as shown in Figure~\ref{fig:The UX interface.}, that automatically generates and recommends design resources based on user intent conveyed via natural language. The key idea is to harness the cutting-edge power of LLMs in managing and activating third-party tools to innovate the design workflow of information graphics with a ChatGPT-like design assistant. The new user interface combines two separate interactive approaches: a \textit{Textual Conversational Interface} enhanced by a tool-augmented LLM and a traditional \textit{Graphical Manipulation Interface}. This integration covers the entire information graphics design process, encompassing tasks such as collecting and generating raw resources and integrating these resources, as depicted in Figure~\ref{fig:teaser}. Users, especially non-professionals, can chat with the intelligent agent in natural language to get nearly completed information graphics, which require only a few final steps of post-editing. 

The challenge lies in effectively translating user intentions, expressed through natural language, into actionable commands that can be executed using third-party tools. To tackle this challenge, we start by identifying a range of common infographic design tasks outlined in existing literature and systems (Section~\ref{sec:study}). These tasks encompass activities such as thematic image design, content collection, etc. Then, we align and configure advanced AI tools to automate these design tasks in Section~\ref{sec:Implementation}. With this set of automatic operations, our LLM tool functions as a \textit{domain-specific scheduler and executor}, which assists users in accomplishing design tasks via a text-based interface. When conversing with users, the LLM dynamically determines which functions, if any, to invoke and automatically initiates the corresponding actions, or continues the conversation as usual. In our LLM-centric interface, a conventional graphical manipulation view is also integrated to support those tasks not aligned with automatic AI methods. Our tool is intended for non-professional users, however, it can also be used for quick design prototyping for professional designers. 

\cmark{Being among the pioneering applications that harness the capabilities of large language models in graphic design, We evaluated GraphiMind conducting a user study to assess its effectiveness and to better understand how users utilize a text-based interface in information graphics design. Through the collection and analysis of user interaction logs, we unveil insights into how GraphiMind seamlessly integrates textual conversations into the design workflow with conventional graphical manipulation tools. The results affirm that our large language model-centric tool manages to alleviate the complexity encountered by non-professional users while supporting their creative capabilities throughout the design process.}

The contributions of this work are:
\vspace*{-5pt}
\begin{itemize}
    \item Identify tasks in information graphics design and align them with state-of-the-art AI technologies and models. These tools can meet the needs of novice users for the corresponding design tasks. 
    \item Propose an LLM-centric interface with the agent \textit{GraphiMind} for information graphics design. Users communicate design needs to the agent via natural language. The agent sources both information and design resources, empowering users to arrange them on a canvas and craft compelling information graphics. 
    \item \cmark{Evaluate GraphiMind, comparing interaction behavior to a baseline system and examining how novice users blend text-based and graphic-based manipulations. We show that GraphiMind significantly shortens the information graphics design process, streamlining the design flow from the content collection to the fine-grained manipulation of objects.}
\end{itemize}
\section{Related Work}\label{sec:rw}

In this section, we first introduce and discuss the related work for information graphic authoring, then cover the literature related to tool-augmented large language models.

\subsection{Information Graphic Authoring}

There is a growing interest in graphics authoring systems that facilitate the creation of expressive graphic designs~\cite{amini2016authoring, satyanarayan2019critical}. 
While commercial tools such as Adobe Illustrator~\cite{adobeillustrator} provide powerful editing capabilities, they typically require a non-trivial degree of design expertise to yield visually captivating results. In contrast, certain research prototype systems prioritize the creative process over offering highly flexible editing functions.
Park et al.~\cite{park2018graphoto} presented Graphoto as a framework that can automatically select a mountain photo that matches the input line graph, and superimpose the line graphs of input data over the mountains. 
Later, Coelho et al.~\cite{coelho2020infomages} continued the idea of embedding commonly used charts into thematic images, and presented a design tool that links web-scale thematic image search with a set of graphics filling and editing functions. 
Kim et al.~\cite{kim2016data} introduced a data-driven graphic design approach that facilitates user interaction by connecting three core encoding channels, i.e., area, length, and position, in vector graphics with data.
In a subsequent study, building on the concept of \textit{lazy data binding}, Liu et al.~\cite{liu2018data} created \textit{Data Illustrator}, which provides increased flexibility in manipulating vector elements.
Several other graphic authoring systems support customizing visual marks with visual embellishments. For example, 
\textit{InfoNice}~\cite{wang2018infonice} supports interactive mark customization in charts. 
Zhang et al~\cite{zhang2020dataquilt} present DataQuilt that allows authors to crop visual and stylistic elements from raster images and re-purpose them to create custom, pictorial visualizations. 
Tyagi et al.~\cite{Anjul2022_info} present \textit{Infographic Wizard}, which takes user-input text in a bullet point form and suggests possible infographics designs with various visual information flows. 

Some authoring systems are more tailored to specific chart types of infographics. For instance, 
Cui et al.~\cite{cui2019text} present the automatic generation of proportion-related infographics given
natural language statements. 
Chen et al.~\cite{chen2019towards} propose an automatic approach to extract the template from a timeline bitmap, and make it extensible for new data.
Other works provide simplified editing techniques for animated graphics. For example, 
Wang et al.~\cite{wang2021animated} propose \textit{InfoMotion}, which automatically infers structures of static infographics and creates an animated version. 
Kazi et al.\cite{kazi2014draco} propose a sketch-based interface, \textit{Draco}, 
which supports users to easily add continuous animation effects to bring static illustrations to life. 
Numerous graphic design authoring systems and tools, such as Adobe Photoshop\footnote{\url{https://www.adobe.com/products/photoshop.html}} and Adobe Illustrator\footnote{\url{https://www.adobe.com/products/illustrator.html}}, have been proposed, with the majority following the creative graphics philosophy of direct graphic manipulation. In contrast, our work harnesses the capabilities of Large Language Models (LLM) and advancements in Artificial Intelligence to provide an end-to-end solution—from data collection to final design—promoting a text-based interface for the creation of information graphics. Further detail on this approach is provided in Section~\ref{sec:study}.

\subsection{Tool-augmented Large Language Models}

Recently, the remarkable capabilities of Large Language Models (LLMs) have garnered substantial attention, leading researchers to investigate their task plan and tool usage potential~\cite{parisi2022talm, mialon2023augmented, qin2023tool, ruan2023tptu}, leveraging the world knowledge from web-scale training data\cite{bommasani2021opportunities}.
Tool-augmented LLMs~\cite{parisi2022talm} refer to a category of LLMs that can automatically invoke the relevant external tools based on user intent, achieved either through fine-tuning~\cite{schick2023toolformer} or by injecting tool-related information via in-context prompts~\cite{liang2023taskmatrix, gupta2023visual, hsieh2023tool}.
One line of research focuses on empowering LLMs to interact with the physical world~\cite{liang2023code, ahn2022can, yao2022react, huang2022language, driess2023palm}.
Another line of research aims to integrate LLMs with various software applications, enable LLMs to tackle tasks, with these models automatically invoking the necessary external tools as needed.
Data-Copilot~\cite{zhang2023data} is an LLM-based system that bridges billions of data with diverse user requests to construct automated data science workflows for various data-related tasks.
TableGPT~\cite{zha2023tablegpt} is a unified framework fine-tuned to allow LLMs to comprehend and manipulate tables using external functional commands.
SheetCopilot~\cite{li2023sheetcopilot} design a state machine-based task planning framework for LLMs to robustly interact with spreadsheets.
VisProg~\cite{gupta2023visual} is a neuro-symbolic method to handle compositional visual tasks given natural language instructions. Visual ChatGPT~\cite{wu2023visual} combines ChatGPT with visual foundation models to solve complex visual tasks using the prompt manager. 
Compared with these works, our framework emphasizes interactivity and user control, rather than solely relying on LLM through natural language instruction to manage the entire workflow, which is not appropriate for information graphics design.
The work most similar to ours is InternGPT~\cite{liu2023internchat}, but it is oriented towards general visual tasks and not specifically towards information graphics design.
To the best of our knowledge, our framework is the first LLM-assisted generative interface for information graphics design.

\section{Identifying Information Graphics Design Tasks}\label{sec:study}

To explore the task space of infographic design, our initial step involved an exhaustive examination of existing literature and systems focused on information graphics design. The primary objective was to acquire a deeper understanding of the inherent design workflow and to precisely identify the core design tasks that are integral to this process. Below, we outline the process through which we compiled the literature corpus and detail the methodology employed for analysis. This led us to identify six essential categories of design tasks within the workflow of information graphic design, which we enumerate in Section~\ref{sec:tasks}.

\subsection{Corpus}

Information graphics encompass a diverse array of visual representations~\cite{smiciklas2012power}. There are two emerging threads found in the literature on information graphics design. One thread centers on \textit{generic composite infographics}, which involve creating visual representations by combining various elements such as icons, text, and images to convey complex information in a visually pleasing manner, as exemplified in studies~\cite{Anjul2022_info, lu2020exploring, s2022content, shi2023reverse, zhang2020dataquilt}. The other thread focuses on \textit{chart-specific infographics}, which manipulates the visual channels of graphic elements in embellished charts to encode information, such as timeline~\cite{chen2019towards}, proportion-based infographics~\cite{qian2020retrieve,cui2019text}, bar charts~\cite{wang2018infonice}, etc.
The scope of information graphics in this work is aligned with the generic composite infographics, i.e., the general compilation of information and graphics. Some open-sourced infographics datasets like InfoVIF~\cite{lu2020exploring} and the dataset contributed by Shi et al.~\cite{shi2023reverse} showcase a substantial collection of composite infographics instances found in \textit{PowerPoint} presentations and knowledge-oriented posters, which hold widespread relevance in daily life.

Having established a clear definition for information graphics, 
we collected the corpus of literature and systems on informational graphic design by snowballing methodology~\cite{wohlin2014guidelines}. First, a core literature set is initialized by searching publications on the topic of information graphics, specifically on their creation technology or systems, from prominent VIS and CHI venues (ACM CHI, IEEE TVCG, and EuroVIS) in recent years (2020-2023). Five core papers are collected~\cite{Anjul2022_info, zhang2020dataquilt, lu2020exploring, coelho2020infomages, shi2023stijl}. Then starting with those papers, we expanded our corpus by including their cited or citing papers on the topic related to `information graphics authoring'. In the end, eleven papers were assembled as the corpus~\cite{cui2019text, kim2019dataselfie, wang2018infonice, kim2016data, chen2019towards, park2018graphoto}. Note that commercial products such as Adobe Illustrator~\cite{adobeillustrator} that support designers in authoring information graphics from scratch with powerful editing functions are very different from our focus and, therefore not included in our corpus. 

\subsection{Methodology}

We analyze these papers to understand the essential tasks in the design workflow of information graphics.
First, we went through all of the papers to get familiar with existing methods and systems, guided by the questions \textit{what are the design tasks and sub-tasks involved in information graphics design}.
Extended from four key infographics design components identified in previous infographic studies~\cite{Anjul2022_info} (\textit{visual information flow layouts}, \textit{visual group design}, \textit{pivot graphics} and \textit{connnecting elements}), an initial set of codes for design operations was generated from information collection to final design output. Each code describes a design action, e.g., `information searching', `icon design', or 'theme figure creation', etc. Then we went through all the papers again to highlight the functions that matched these codes. New codes were allowed to be added if needed. Then codes with common patterns were identified and combined into a theme of the design task. We returned to the data set and reviewed themes to adjust them. Finally, six essential themes of design tasks are identified.

\subsection{Design Tasks in Information Graphics}
\label{sec:tasks}

Six design tasks are distilled from information graphic design. In this part, we introduce these tasks and describe how they are supported by papers in the corpus.  

\paragraph{Task 1: Information Collection}

\begin{wrapfigure}{lth!}{0.06\textwidth}
  \centering
  \includegraphics[width=0.08\textwidth]{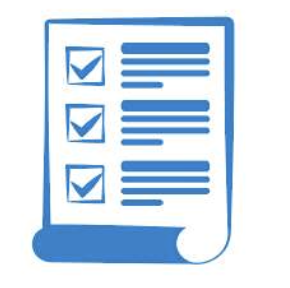}
\end{wrapfigure}

The theme of Information Collection focuses on the gathering, selection, and arrangement of relevant data and content for the infographic. Information collection is the foundation upon which the entire infographic is built. This involves the meticulous process of curating information to effectively convey the intended message. \cmark{In many cases, information is collected beforehand, separate from the graphic authoring activities. For instance, in Infographics Wizard~\cite{Anjul2022_info}, users specify content in a markdown format, while in another case, users load a data file as demonstrated in~\cite{coelho2020infomages}.}

\paragraph{Task 2: Visual Element Design}

\begin{wrapfigure}{lth!}{0.06\textwidth}
  \centering
  \includegraphics[width=0.08\textwidth]{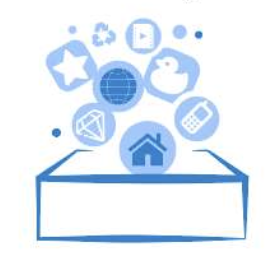}
\end{wrapfigure}

Visual Element Design encompasses the creation and integration of individual graphical components within the infographic. This theme involves decisions related to icons, illustrations, charts, graphs, and other visual elements that visually represent the data and concepts. Most of the visual element design aspects are heavily reliant on user-generated content or customization, which can introduce additional complexities and time-consuming processes. \cmark{Graphic authoring systems offer support for utilizing existing artworks through various means. This includes selecting from a library~\cite{kim2016data}, uploading by users~\cite{wang2018infonice}, or extracting from existing designs~\cite{zhang2020dataquilt}. Additionally, some systems also allow users to sketch visual elements from scratch ~\cite{kim2019dataselfie}.}


\begin{wrapfigure}{lth!}{0.06\textwidth}
  \centering
  \includegraphics[width=0.08\textwidth]{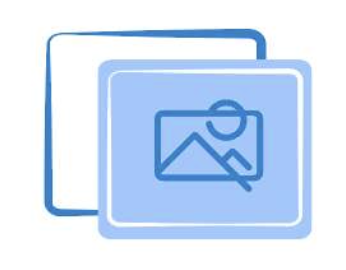}
\end{wrapfigure}

\paragraph{Task 3: Pivot Figure Design}
Pivot Figure, or thematic images~\cite{coelho2020infomages}, refers to the central or pivotal visual element that encapsulates the main message or focal point of the infographic. This figure serves as an anchor, around which other elements revolve. The design of the pivot figure should be engaging and attention-grabbing to draw viewers into the content. This includes mastery of advanced graphic design software, an understanding of color theory, and typography, and an ability to convey complex ideas through visual metaphors. \cmark{Advanced design skills are typically necessary for creating a pivot figure, leading many graphic design systems to incorporate external resources for support. This includes enabling image searches~\cite{coelho2020infomages}, suggesting figures from datasets~\cite{park2018graphoto}, allowing users to upload their images~\cite{Anjul2022_info}, or change color palettes suggested by themes~\cite{shi2023stijl}.}

\begin{wrapfigure}{lth!}{0.06\textwidth}
  \centering
  \includegraphics[width=0.08\textwidth]{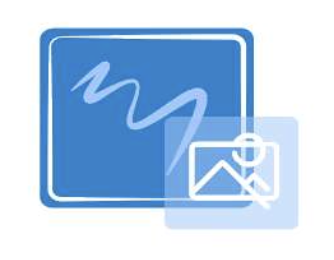}
\end{wrapfigure}

\paragraph{Task 4: Background Design}

Background Design encompasses the visual backdrop against which the entire infographic is presented. While often subtle, the background plays a crucial role in setting the mood, enhancing contrast, and ensuring the readability of the content. The choice of colors, patterns, and textures in the background affects the overall aesthetic. \cmark{Editing functions, such as adjusting color and texture, are commonly provided to facilitate fundamental background design. Some graphic design systems also allow users to upload a background image~\cite{Anjul2022_info}.} Unlike basic backgrounds, integrating sophisticated background designs into infographics requires a deeper understanding of graphic design principles, image manipulation techniques, and the ability to seamlessly blend thematic elements.

\begin{figure*}[tbp]
    \centering    \includegraphics[width=\textwidth]{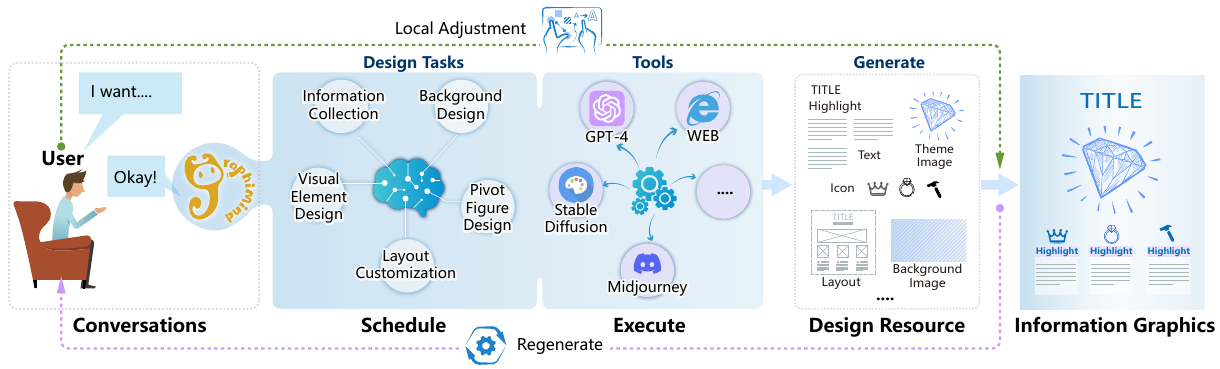}
    \caption{\cmark{GraphiMind Workflow: it combines conversational interactions and graphical manipulation for infographic creation. Users communicate with an intelligent agent to initiate design tasks, which are executed using tools like Stable Diffusion. The design resources gained can be manipulated on a canvas. This process allows for an iterative blend of AI-driven automated resource generation and user-driven adjustment.}}
    \label{fig:workflow}
\end{figure*}

\paragraph{Task 5: Layout Customization}

\begin{wrapfigure}{lth!}{0.06\textwidth}
  \centering
  \includegraphics[width=0.08\textwidth]{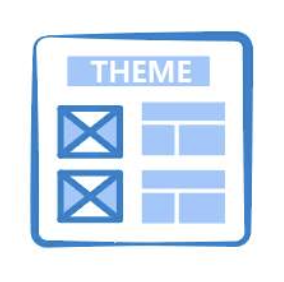}
\end{wrapfigure}

Layout Customization involves arranging all elements within the infographic to convey the visual information flow~\cite{lu2020exploring}. This theme includes decisions about the hierarchy of content, the order of sections, and the strategic placement of key points to optimize information absorption and retention. Effective grouping and connections in an infographic enhance its readability and help convey complex ideas more intuitively. \cmark{In layout design, some approaches involve extracting and repurposing layouts from existing infographics, including those based on timelines~\cite{chen2019towards} and proportion-related graphics~\cite{cui2019text}. Alternatively, some methods search for layouts based on user-provided sketches and offer recommendations~\cite{lu2020exploring, Anjul2022_info}.}

\paragraph{Task 6: Local Adjustment}

\begin{wrapfigure}{lth!}{0.06\textwidth}
  \centering
  \includegraphics[width=0.08\textwidth]{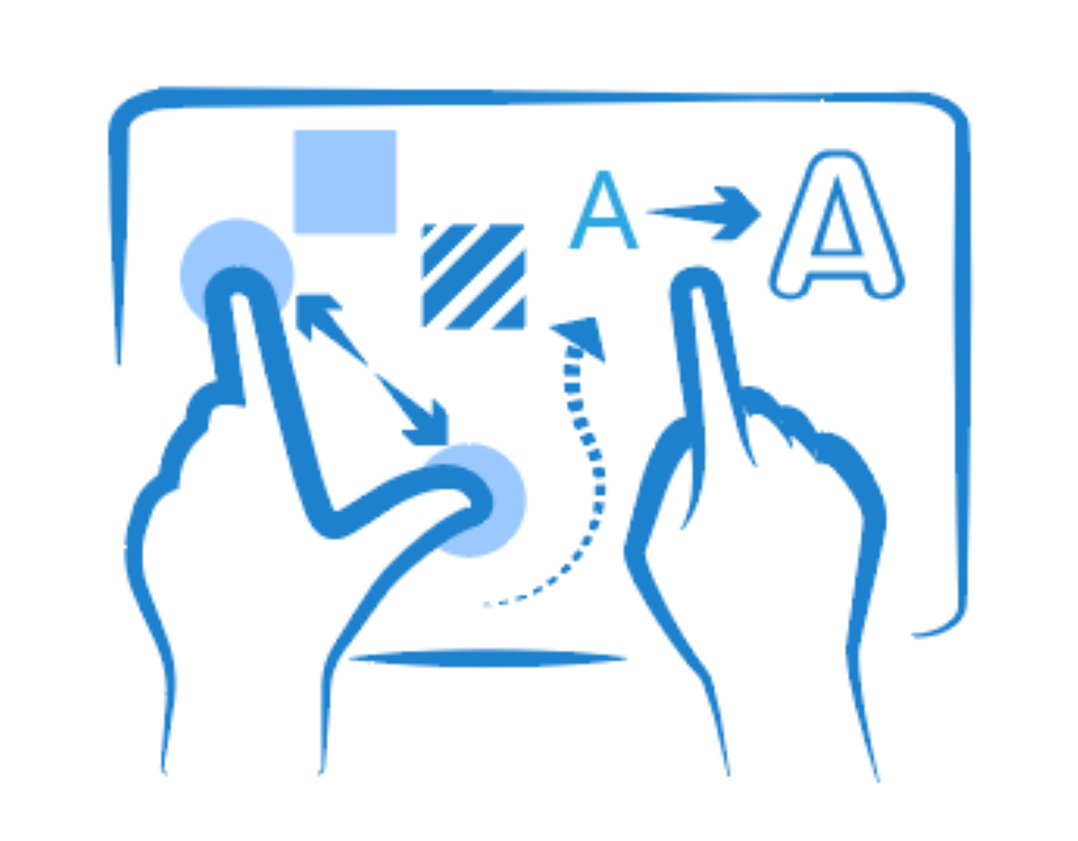}
\end{wrapfigure}

Local adjustment refers to the visual appearance modification of design assets individually, including altering the position, size, color, font style, font size, or other visual attributes of elements to achieve a desired effect or enhance the overall aesthetics. \cmark{It is a task commonly performed in design, image editing, or any context where precise and targeted modifications to visual elements are necessary. Many graphic design systems widely support direct manipulations on the canvas, encompassing actions like dragging, resizing, and applying diverse visual effects such as grayscale and transparency~\cite{Anjul2022_info, coelho2020infomages}.}

\begin{table*}[tbp]
\centering
\resizebox{\textwidth}{!}{%
\begin{tabular}{ccc}
\toprule
           Design Task & Tool & Is Managed by \\
\midrule
\makecell{Information Collection \\ \& Visual Element Design} &  \makecell{The ChatGPT prompted to simultaneously generate a title and bullet points on a design topic.\\ Each point includes an icon keyword (convertible to SVG via API), a headline, and content.} & \multirow{5}{*}{Agent} \\
\cline{1-2}
   Pivot Figure Design &    The Stable Diffusion prompted for pivot figure. &  \\
     Background Design &    The Stable Diffusion prompted for background figure. &  \\
  Layout Customization &    The GPT-4 prompted to design layout through domain-specific language. &  \\
\midrule
      Local Adjustment &    \makecell{Drawing board features include options for dragging and resizing, \\adjusting icon color and line thickness, and setting text font and size, among others.} & Canvas \\
\bottomrule
\end{tabular}
}
\caption{\cmark{The Correspondence between Design Tasks and System Implementations in GraphiMind: 
Agent refers to a tool-augmented large language model within the \tci, and Canvas refers to the \gmi. Each of them is responsible for certain system functionalities.}}
\label{tab:correspondence}
\end{table*}

\section{Overview}

\begin{figure*}[tbp]
    \centering    \includegraphics[width=0.88\textwidth]{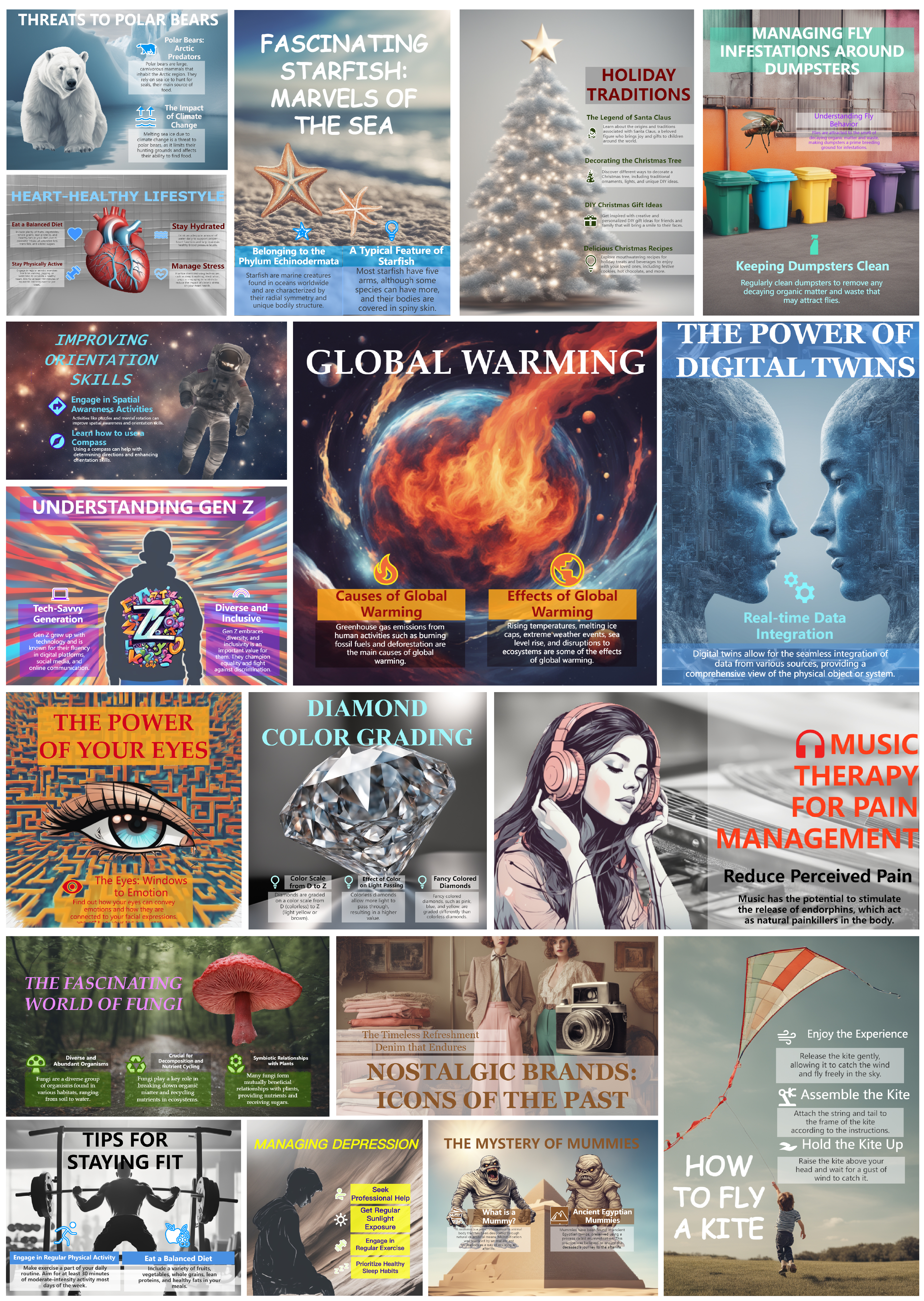}
    \caption{\cmark{The Gallery of Information Graphics Created Using GraphiMind: it demonstrates a diverse range of information designs, encompassing various topics, contents, imagery, layouts, and so on.}}
    \label{fig:gallery}
\end{figure*}

\cmark{
The interface of \gm, as shown in Figure~\ref{fig:The UX interface.}, comprises two side-by-side sections: on the left, the \tci , where users can chat with an intelligent agent, and on the right, the \gmi, serving as a canvas.
The dynamic workflow of GraphiMind, depicted in Figure 3, represents a novel design experience that blends text-based and graphic-based interactions and results in the creation of an infographic, fulfilling the design tasks specified in Section 3.3. This section provides a detailed explanation of how \gm~ manages this hybrid workflow. The subsequent section will delve into the specifics of the system's implementation.

The workflow of \gm~, as shown in Figure~\ref{fig:workflow}, involves a cyclical process between the two interaction modes: users interact with the agent to gather design resources, and users manipulate these resources on the canvas. In the dialogue between the agent and the user, once the user submits a message, the agent determines whether to schedule a design task or to continue with a regular conversation. This decision is based on the content of the received message and the predefined purpose and significance of each design task. When no design task is required, the agent functions as a conversational chatbot, engaging in normal dialogue with the user.
Conversely, when a design task is essential, the process transitions into the \textit{Scheduling} phase. The agent then identifies the appropriate tool and arguments by inferring users' intentions from the conversation. For example, in the design task "pivot figure design", the agent schedules the Stable Diffusion tool, using an image caption that aligns with the user's intent as inferred from the dialogue context, such as "a cute cat".
The chosen tool and its input parameters are then forwarded to the subsequent \textit{Execution} phase. The specific execution of these tools varies, ranging from local instances of AI models (e.g., Stable Diffusion) to APIs of search engines. Detailed descriptions of these implementations are provided in Section~\ref{sec:Tool Library}.

Upon completion of each execution process, relevant design resource will be returned. Tailored to the specific design task (section~\ref{sec:tasks}), 
these resources vary in formats, such as text for information collection, SVG for visual elements, and PNG for pivot figure and background design. As shown in Fig~\ref{fig:The UX interface.}, these resources are seamlessly embedded within ongoing dialogues, in which the resources can be further processed, such as segmentation.  
To enhance the usability of the canvas, GraphiMind automatically positions these resources on the canvas and also offers flexibility for users to manually place them via drag-and-drop, catering to individual preferences. Also, all design resources on the canvas are initially assigned default visual properties, such as font size, colors, etc. Users can edit these properties using the toolbar below the canvas (the bottom right in Figure~\ref{fig:The UX interface.}). 
This manipulation (referred to as task 6 'local adjustment' in section \ref{sec:tasks}) includes adjusting font size, altering colors, repositioning, etc.

GraphiMind enables users to seamlessly transition between textual conversations and canvas manipulation. In cases where the design falls short of user expectations, they have the option to regenerate new design resources. This entails rejoining the ongoing dialogue with the agent, and revisiting the scheduling and execution processes until the desired design is achieved. Figure~\ref{fig:gallery} shows a gallery of infographics crafted using GraphiMind.
}

\section{Implementation}
\label{sec:Implementation}

\cmark{
This section details how GraphiMind manages and implements the design tasks, with the corresponding relationships depicted in Table ~\ref{tab:correspondence}.
Specifically, GraphiMind is composed of three fundamental components, i.e., \textit{Agent} engineered to identify relevant design tasks based on user intent and subsequently activate related tools, \textit{Agent-managed Tool Library} assembled to accomplish a spectrum of design tasks, and \textit{Canvas} tailored to facilitate design tasks that are not amenable for natural language execution. Further details are available in the Appendix.
}

\begin{figure*}[tbp]
    \centering    \includegraphics[width=\textwidth]{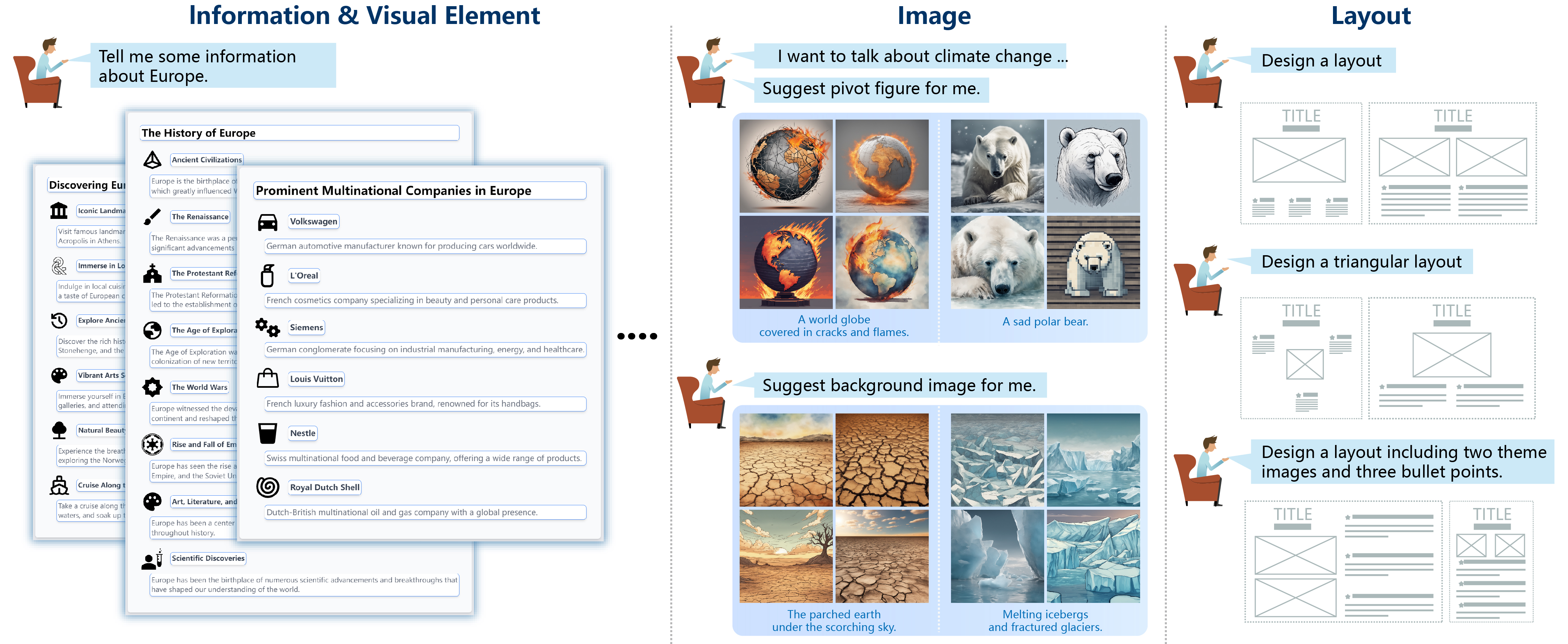}
    \caption{\cmark{The Diverse Design Assets by GraphiMind in the Dialogue between Agent and User: the diversity stems from the variety in the agent's reasoning processes as well as the inherent diversity of the tools themselves.}}
    \label{fig:diversity}
\end{figure*}

\subsection{Agent}

\cmark{In GraphiMind, the agent's responsibility is to reason and invoke the relevant tools from a predefined set of tools based on the user's intent expressed in natural language.}
Among various methods for creating a centralized tool management agent using LLM, ChatGPT distinguishes itself through its ease of use and outstanding performance, due to the function calling feature.
While our study focuses on utilizing ChatGPT as the tool manager of GraphiMind, it should be noted that the GraphiMind framework is generalized and can be adapted to integrate other tool-augmented LLMs.
\cmark{
Following the official guidelines\footnote{\url{https://platform.openai.com/docs/guides/function-calling}} provided by OpenAI, it is essential to customize a comprehensive tool signature for each item within the predefined toolset. This includes detailing the names and descriptions of each tool along with its input arguments. We outline the methodology adopted in GraphiMind for crafting the signatures for each tool.
}

A tool signature encodes two essential types of information to successfully guide the agent: explanatory and exemplary information.
Explanatory information is principally derived from the name and description tied to the tool and its input parameters, offering a foundational understanding of the tool's intended purpose.
In addition, exemplary information provides context-specific insights or illustrative examples, demonstrating the descriptions associated with each tool and its input parameters.
\cmark{We take the tool signatures for the tasks of "pivot figure design" and "background design" as an example, to explain the explanatory and exemplary concepts (more implementation details shall be explained in section~\ref{sec:Pivot Figure and Background Design}). Both design tasks of pivot images and background images fall into the category of image generation but with different focuses, thus requiring two separate tools to handle these design tasks. To enable the agent to understand and differentiate between these two tools, it is necessary to incorporate distinct explanatory and exemplary information in their tool signatures. Specifically, for the tool responsible for pivot figure design, its setting in GraphiMind focuses on a central object and its accompanying information, exemplified by "a smiling dog wearing a hat". Conversely, the tool designed for background image tasks in GraphiMind emphasizes a descriptive environment or background scene, illustrated by the example of "the park under warm sunlight".
The rationale behind these descriptions is to help the agent reason the user's intention of designing pivot figures or background images based on context. For instance, when the user's conversation theme revolves around polar bears and climate change, the GraphiMind agent might deduce that "a crying polar bear" would serve as an apt pivot figure, and "the melting iceberg" as a suitable background image. This intelligent reasoning not only alleviates the cognitive load for the user but also lowers the barrier to design.
}

By adhering to both forms of explanatory and exemplary information, agent is able to accurately call appropriate tools and reason the input arguments for each tool. This results in better alignment with specialized task conditions and reduces the potential for contextual misunderstandings.

\subsection{Agent-Managed Tool Library}
\label{sec:Tool Library}

\cmark{
In this section, we elaborate on the tools integrated in GraphiMind. In order to provide a better understanding, we describe tool implementation according to the associated design tasks. The design assets produced in accomplishing these tasks are diverse, as shown in Figure~\ref{fig:diversity}, benefiting from the varied reasoning capabilities of the agent and the diversity inherent in each tool.

Additionally, we implemented two more tools to achieve auxiliary tasks (details in Section~\ref{sec:Auxiliary Tasks}), to enhance the user experience. Apart from the image clipping tool in the auxiliary tasks, all the tools involved in this section are uniformly managed by the agent, i.e., they can only be activated and executed through dialogues with the agent. It is worth mentioning that our agent-managed tool library exhibits a high degree of scalability and flexibility, in which each tool can be upgraded independently without affecting the overall integrity of the system.
}

\begin{figure*}[tbp]
    \centering
    \includegraphics[width=\textwidth]{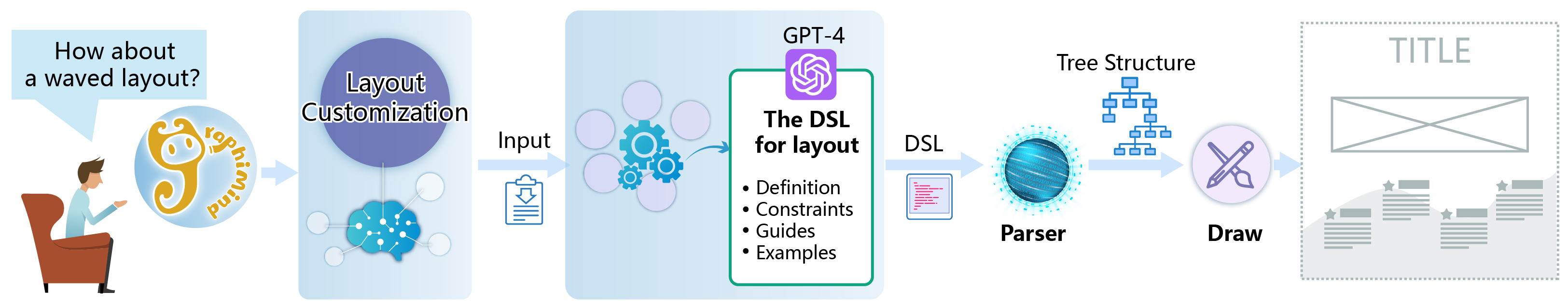}
    \caption{\cmark{The Pipeline of Layout Customization Tool: this process encompasses user-agent interaction, GPT-4 activation, followed by parsing and drawing stages, culminating in the final layout generation.}}
    \label{fig:Implement - Layout Customization.}
\end{figure*}

\subsubsection{Pivot Figure and Background Design}
\label{sec:Pivot Figure and Background Design}
\cmark{The tool used to execute these two design tasks is} Stable Diffusion XL 1.0 \footnote{\url{https://huggingface.co/stabilityai/stable-diffusion-xl-base-1.0}}, a powerful open-source generative model developed by Stability AI \footnote{\url{https://stability.ai/}}. We configure it to produce two distinct types of images, the pivot figures and background images, achieving by specialized input caption templates. For example, the template "a background of \{ \}" generates background images, whereas "a focused image of \{ \}" is used for pivot images. In conversation, the agent autonomously reasons between generating a pivot figure or background image based on the context of the dialogue.
In addition, besides the input caption, we have also add two additional arguments: style (e.g., watercolor, 3D render), and effect (e.g., blur, focused) to help identify the diverse needs of users. Technically, the tool relies on Hugging Face's diffusers library in PyTorch framework \footnote{\url{https://pytorch.org/}}. The image generation process consists of a 50-step diffusion sequence and takes approximately 12 seconds per image when executed on a GTX 3090Ti GPU. A unique feature of Stable Diffusion is its ability to produce diverse images from identical input prompts, thereby enabling the concurrent operation of multiple instances without compromising image diversity.

\subsubsection{Information Collection and Visual Element Design}
\cmark{
We utilize a single tool, rather than employing two distinct tools for the concurrent execution of information collection and visual element design. This approach is predicated on the intrinsic correlation between these two design tasks: it ensures that the visual element and its content within a bullet point are semantically aligned. To achieve this, the tool is specifically composed of two sequentially connected components.
}

The first component is for information collection, and we employ ChatGPT through prompt engineering. Due to ChatGPT's inherent rich world knowledge and advanced natural language understanding, we can customize it through prompt engineering to specifically collect information for infographic design.
This instance of ChatGPT is specifically configured to fulfill two primary objectives. 
The first goal centers on creating a complete array of related information elements, which encompass the title of the information graphic and bullet points that may include featured icons, headlines, and the main content. The second objective pivots towards structuring the gathered information and the aim here is to format them into a JSON object, thereby simplifying data parsing and facilitating their seamless integration into the canvas.
As a result of these two objectives, the return by ChatGPT comprises two primary keys "title" and "bullet points", ready for parsing as a JSON object. 
Specifically, the infographic topic encapsulates the overall subject matter of the infographic.
On the other hand, the bullet points are organized as an array.
Each entry within this array contains three subsidiary keys: "icon keyword," "headline," and "content."
These subsidiary keys serve distinct but interconnected purposes:
the icon keyword is a word or phrase used for retrieving appropriate icons, facilitating ease of access;
the headline offers a succinct highlight of each bullet point's central theme, providing a quick overview;
the content, in alignment with the current headline, furnishes detailed information.

\cmark{
The second component is dedicated to the design of visual elements. Specifically, this involves transforming icon keywords, gathered in the information collection step, into icons in SVG format.
}
We utilize the Iconify database\footnote{\url{https://iconify.design/}}, a continuously updated resource offering a diverse array of SVG icons. Through the search API provided by Iconify, each icon keyword is converted into a corresponding SVG icon, which is instantly viewable and can be adjusted on the canvas with properties such as color and thickness. Importantly, this search process is scalable; it can synchronize with updates in the Iconify library or incorporate additional online icon repositories. Furthermore, it supports user customization, enabling users to create a personalized icon database. This feature allows for the tailoring of icons to meet specific thematic or stylistic preferences.

\subsubsection{Layout Customization}
\label{sec:Layout Customization}
\cmark{
The layout customization tool generates layouts based on design intentions expressed in natural language, as the agent interacts with the user through natural language.
}
Most information graphics are found with hierarchical layouts, i.e., extending from a global macro-level layout to micro-level clusters of interrelated visual elements and groups~\cite{lu2020exploring, shi2023reverse}. In light of GPT-4's strong capabilities in structural reasoning, we are inspired to explore its potential in layout design. However, a challenge emerges when applying GPT-4 to layout design: how to constrain and transform the textual conversations with GPT-4 to a visual layout.


To tackle this challenge, we designed a Domain-Specific Language (DSL) for constructing information graphics layouts. The DSL adopts a tuple-based hierarchical structure with two primary types of nodes: coordinate nodes and container nodes. Coordinate nodes serve the purpose of establishing a recursive and hierarchical framework for the layout, essentially acting as scaffolding that structures the information graphics. Container nodes, on the other hand, encapsulate specific design elements like titles, images, or icons that populate the layout.
To manage complexity while providing a structured alignment of design elements, our domain-specific language features two critical aspects: constraints and guides. Constraints refer to layout generation rules that aim to create a visually harmonious interface, for instance, minimizing significant overlaps between different design elements. On the other hand, guides refer to specific directives in information graphics design, such as allowing only one icon, one headline, and one content element within a container group. By integrating both constraints and guides, our DSL serves as a comprehensive framework for automating and optimizing layouts in information graphics design.

\cmark{
Practically, the layout customization tool was developed by instantiating GPT-4 with the integration of the aforementioned DSL document into its context.
}
Figure~\ref{fig:Implement - Layout Customization.} illustrates the scheduling and execution processes of this tool during the interaction between the user and the agent.
Initially, the agent discerns the user's intent for layout customization from the user's message and proceeds to activate GPT-4, guided by the domain-specific language document. This process involves supplying an input argument - a string with the directive such as "Generate a waved layout". After this activation, strings in the DSL representation of generated layouts are generated and returned in the execution phase. Then the strings go through a Parser and a Draw module. The parser's role is to decode the DSL representation, translating it into a corresponding tree structure. The draw module effectively renders the tree structure into a tangible layout. This layout is then transformable into an editable format on the canvas, ready for further manipulation or use.
The flexibility of our DSL makes it amenable to extensions that can accommodate various application scenarios or more complex design requirements. As advancements in language language models continue to develop, we anticipate that this tool will benefit from these developments, thereby amplifying its efficacy and range of capabilities.

\subsubsection{Auxiliary Tasks}
\label{sec:Auxiliary Tasks}
\cmark{
To enhance the infographic design process, we implemented two innovative tools, each specifically tailored for distinct auxiliary tasks.
}
The first tool, an image manipulation tool, can be automatically activated by the agent and facilitates image modification using natural language commands. This tool, built on the deep learning model InstructPix2Pix\cite{brooks2023instructpix2pix}, allows users to make a range of edits, for example, altering the style to watercolor or adjusting imagery to depict snowy conditions, without requiring extensive graphic design expertise. \cmark{Figure~\ref{fig:teaser} shows an example of making an image snowy}. It is useful when users wish to edit existing images rather than recreate them entirely.

\cmark{
The second one, a user-friendly image clipping tool, is powered by the deep learning model SAM (Segment Anything Model)\cite{kirillov2023segment}. This tool, operated directly by the user rather than managed by an agent, efficiently streamlines the extraction of specific areas within images. Within the GraphiMind system, this tool offers three methods for selecting areas of interest, by point, line, and rectangle. The point method involves a single click within the area, such as click a dog in an image to extract the entire animal. The line method requires drawing a line inside the desired area, while the rectangle method involves outlining the area with a rectangular frame. SAM then intelligently segments the selected object based on semantic understanding.
}

\subsection{Canvas}
\cmark{
Some design tasks are inherently more intuitive when performed on a canvas interface as opposed to natural language input.
For example, when a user wishes to change the position of an element on canvas, manipulating it directly is obviously more efficient and user-friendly than asking agent to do it by natural language.
In light of this, GraphiMind incorporates a canvas as the final piece in assisting users with the infographics design. The following section introduces a suite of tools designed for local adjustment (the task 6 in section~\ref{sec:tasks}) and the drag-and-drop mechanism establishing a bridge between the agent and the canvas.
}

\subsubsection{Local Adjustment}

\cmark{
In Section 3, local adjustment is defined as the process of modifying the visual appearance of individual design assets. This encompasses alterations to position, size, color, font style, font size, and other visual attributes. It is evident that manipulating these properties directly on the canvas, either by mouse actions or through the selection of tools from the toolbar, is more convenient than controlling them using natural language. All assets on the canvas are amenable to dragging and resizing. Specifically for text assets, color adjustments can be made to the text and mask colors. Additionally, modifications can be applied to the font and size, as well as toggling between bold and italic styles and altering text alignment within the box. Regarding icon assets, adjustments can be made to their fill and edge colors, as well as the thickness of their edges. Given that GraphiMind is essentially a prototype system, our focus has been more on enhancing the \tci~ rather than the canvas functionality. It is important to note that functions on the canvas can be easily extended.
}

\subsubsection{Drag-and-Drop Mechanism}

\begin{figure*}[htbp]
    \centering    \includegraphics[width=\textwidth]{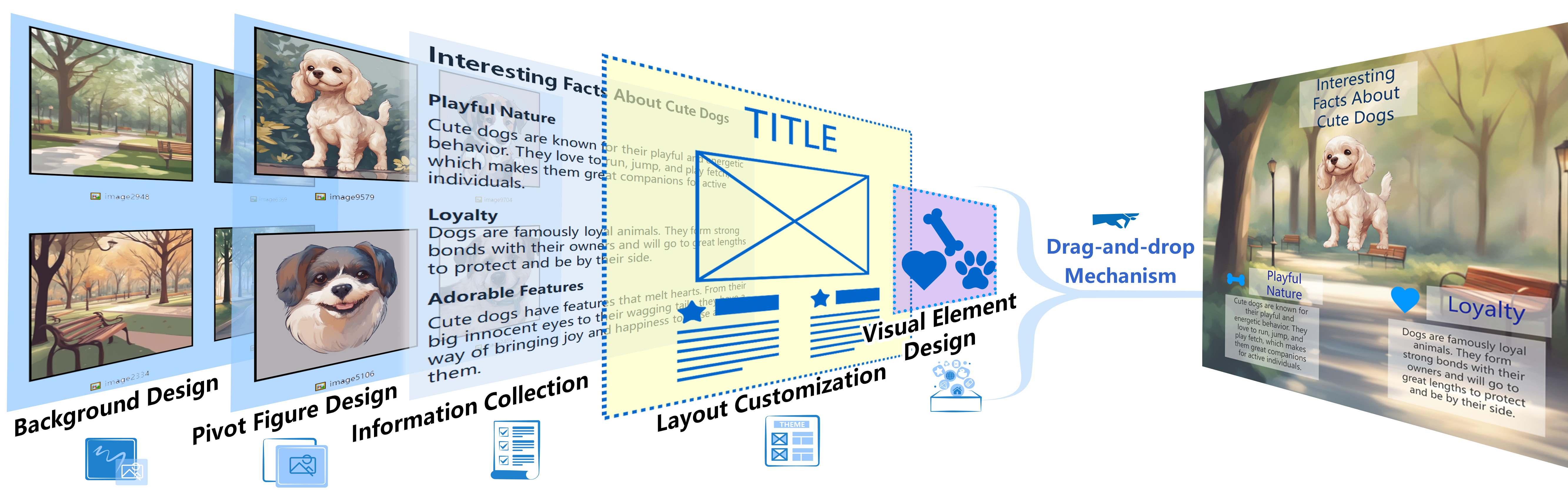}
    \caption{\cmark{The Aggregation of Diverse Design Assets through Drag-and-Drop Mechanism: this method allows users to seamlessly integrate a variety of design assets provided by agents, thereby simplifying and accelerating the process of creating information graphics.}}
    \label{fig:drag-and-drop}
\end{figure*}

\cmark{
This mechanism enables the efficient placement of various design resources onto the canvas, from text blocks to other design elements (shown in Figure~\ref{fig:drag-and-drop}).
This functionality can be executed either automatically by GraphiMind or manually by the user, serving as two complementary modes. The automatic placement is particularly effective during the initial stages of design. For instance, when an agent generates pivot images, GraphiMind will automatically sample and place these images onto appropriate containers (defined in Section~\ref{sec:Layout Customization}) on the canvas. When the agent suggests layouts and displays them, it would be impractical for users to manually arrange these on the canvas. In such cases, GraphiMind facilitates user interaction by allowing a simple click on the suggested layout in the textual conversational interface, which then automatically arranges it on the canvas while adapting the existing design resources to accommodate the new layout. This proactive placement ensures that the canvas is populated with suitable design assets as much as possible, providing quick visual feedback even if it does not fully meet the user’s final vision. Additionally, users also can engage in drag-and-drop manually. This method comes intuitively along with human perception and interaction nature. Especially when placing icons or editing text, a direct touch offers a more natural interaction.
}

\section{Evaluation}

\cmark{
The key idea of GraphiMind is to incorporate the power of LLM, especially its power in task scheduling, to simplify the design workflow of infographic design via a text-based conversational interface. 
Therefore, our main goal was to answer twofold research questions: (1) \textit{how an LLM-centric text-based conversational interface changes the process of infographic design compared to the use of a standard tool}; (2) \textit{how GraphiMind supports creativity and how participants perceive the use of GraphiMind}, highlighting the advantages and problems of an LLM-centric interface as a tool for creating information graphics.

\begin{table*}[tbp]
\centering
\resizebox{\textwidth}{!}{%
\begin{tabular}{rcl}
\toprule
 Index &                    Property &                                The statement shown to participants\\
\midrule
     1 &                  Enjoyment &                           I enjoy using this tool. \\
     2 &                Exploration & Using this tool, I can easily explore many different ideas, options, designs, or outcomes. \\
     3 &             Expressiveness & This tool allows me to design in a very expressive way. \\
     4 &       Results Worth Effort & I am satisfied with the results I get from the tool. \\
     5 &                Ease of Use &     This tool is intuitive and easy to understand. \\
     6 &          Layout Adjustment & Using this tool, I can easily adjust and optimize the layout of infographics. \\
     7 &        Layout Adaptability & The layouts provided by this tool adapt well to different types of content and information quantities. \\
     8 &           Resource Variety & I can obtain a variety of design resources (such as images, icons). \\
     9 &         Resource Relevance & The design resources (such as images, icons) provided by this tool are closely related to my design theme. \\
    10 &              Information Accuracy & The text information provided by this tool accurately expresses the theme and content I want. \\
    11 & Information Collection Efficiency & This tool has improved my efficiency in collecting text information. \\
    \midrule
     Index &                    - &                                Open-ended Question\\
     \midrule
     12 & - & What features in the system do you think help create infographics? \\
     13 & - & What aspects of the system do you think need improvement? \\
     14 & - & What are the pros and cons of using canvas and dialogue interactions in infographic design? \\
\bottomrule
\end{tabular}
}
\caption{\cmark{Overview of the  Questionnaire in the Second Part: the first 11 items were presented as 5-point Likert scale questions, ranging from strongly disagree to strongly agree. The last three questions were open-ended questions that were verbally asked, recorded, and transcribed.}}
\label{tab:questionaire}
\end{table*}

 \subsection{Methodology}

The study included two main parts. In the first part, participants were asked to create two information graphics with either GraphiMind or PowerPoint. In the second part, participants were asked to freely use GraphiMind, after which we surveyed participants.

\subsubsection{Study Design} For the first part of the study, each participant used only one system - either GraphiMind or a baseline system. We used a between-subject design since we wanted to be able to directly compare the use of GraphiMind to the use of a baseline system on the exactly same tasks. We chose PowerPoint as the baseline system considering its wide popularity in information and graphic editing as well as its intuitive and easy-to-use interface. We did not choose a professional graphic tool such as Adobe Illustrator~\cite{adobeillustrator}, since these products have a high learning curve and could overwhelm design novices who are the target users of this work. 
To align the baseline system with the design functions outlined in Section ~\ref{sec:tasks} similar to GraphiMind, the baseline group was permitted to utilize Internet search tools, including Google Images and any other online design resource.
In the subsequent text, we abbreviate the baseline system (i.e., PowerPoint + Internet) as \textit{PowerPoint}. 

In the second part of the study, all participants experienced free-form use of GraphiMind after which they were asked to fill in a questionnaire to ask participants about their general impressions and examine the creative process of GraphiMind.

\subsubsection{Participants} We recruited 16 participants through online advertisements posted at a local university, specifically seeking individuals without professional design expertise. Out of the 16 participants, 11 identified themselves as males and five as females, 6 were undergraduate and 10 were graduate students. Twelve participants (75\%) were aged between 18 and 24, and 4 were between 25 and 35 years old. All of the participants reported minimal-to-none experience in graphic design (6 claimed none and 10 had minimal experience). In terms of familiarity with PowerPoint, three participants reported low proficiency, five medium, and eight relatively high proficiency.
Participants were compensated with a gift equivalent to 10 USD.

\subsubsection{Set-up and Procedure} The 16 participants were randomly allocated into two groups, each comprising eight participants. For the first part of the study, one group only used GraphiMind, while the other group only used PowerPoint. The experiment was carried out on an individual basis, one participant at a time. Each participant was invited to a desktop equipped with a 27-inch display screen, with a mouse and keyboard for their use. In the first part the experiment coordinator (one of the authors) first gave an introduction to information graphics. This included showcasing several examples and explaining the experimental system, i.e., GraphiMind for one group and PowerPoint for the other. All of the functions related to the six infographic design tasks (introduced in Section~\ref{sec:tasks}) were explained and demonstrated in a randomized order, to avoid the implication of any design workflow. Next, participants were asked 
to create two information graphics. The first on the topics of `Your Favorite Character' and the second on the topic of `Ancient Civilizations'. These broad topics were chosen to allow participants the flexibility to use their creativity without being confined by a highly specific or technical subject matter. For each of the two topics, the experiment coordinator first introduced the scope of the topic, and then the participant was instructed to design an infographic they liked on the topic at hand. Participants were explicitly notified that their screens would be video-recorded. 

In the second part of the study, GraphiMind was shown to the participants of the PowerPoint group, and detailed explanations were given about its use. Then, all participants conducted a free-form examination on GraphiMind in which they were asked to freely experiment with the system, try out the system’s functions, and create a graphical design of their choice (for the PowerPoint group). This part was limited to a maximum of 30 minutes.
At the end of this part, participants were asked to fill in a questionnaire  to ask participants about the general impressions and creative process of GraphiMind. The entire session lasted approximately an hour and a half.

\begin{figure*}[tbp]
    \centering    \includegraphics[width=\textwidth]{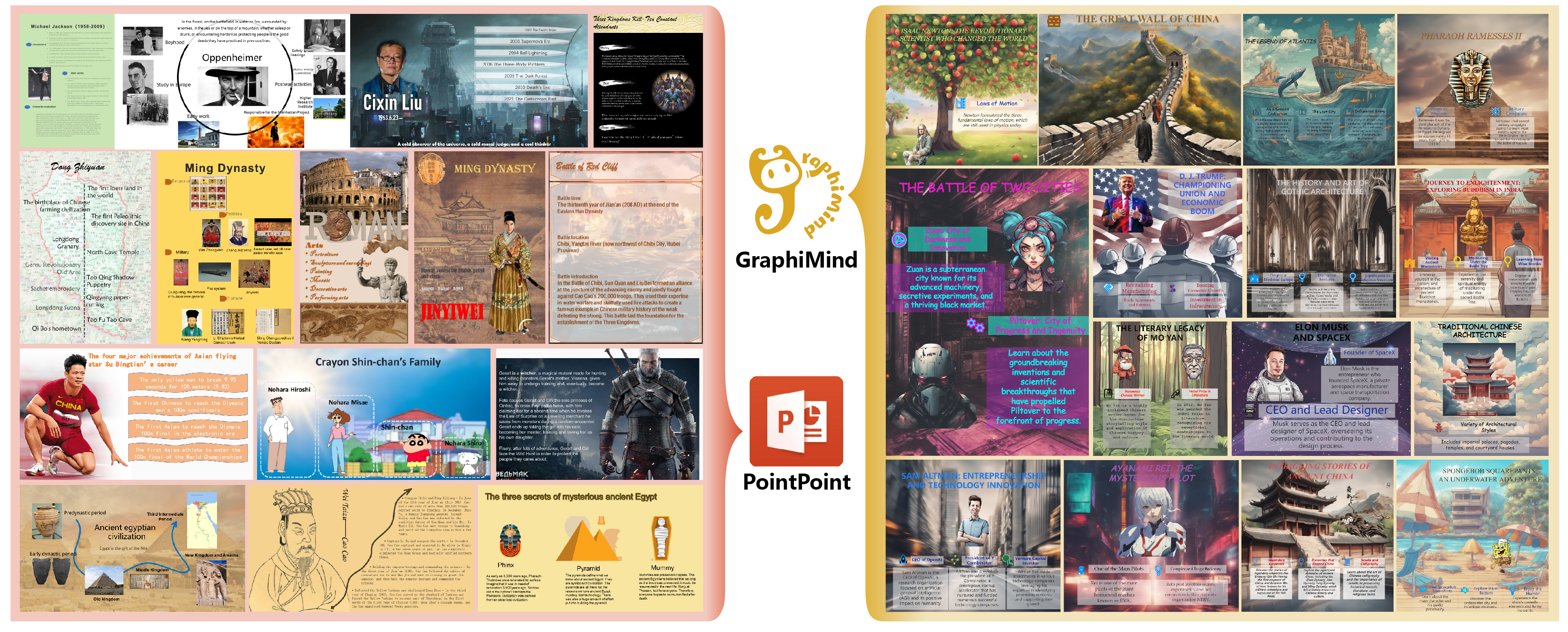}
    \caption{\cmark{Information Graphics Examples Created in the User Study from the GraphiMind Group (right) and the PowerPoint Group (left).}}
    \label{Fig - Visual Quality}
\end{figure*}

\subsubsection{Measures}
For the questionnaire in the second part, we adopted the \textit{Creativity Support Index (CSI)}~\cite{cherry2014quantifying} to measure GraphiMind's creativity support. The original CSI has six dimensions of creativity support, including Exploration, Expressiveness, Immersion, Enjoyment, Results Worth Effort, and Collaboration. Since GraphiMind is not a collaboration software, we used four of the CSI dimensions: Enjoyment, Exploration, Expressiveness, and Results Worth the Effort.
Additionally, we added a question about ease of use and six more questions surveying the extent of creativity supported by specific functions, including layout, resource, and information, each having two questions. Finally, we added three open-ended questions aimed at eliciting comprehensive user feedback on the advantages and disadvantages of GraphiMind, with a particular
focus on users’ perceptions of canvas and dialogue interactions. Overall, the questionnaire included 11 5-point Likert scale questions and 3 open-ended questions, see Table \ref{tab:questionaire}.

\subsubsection{Analysis} Two of the authors worked together to analyze and code the recorded videos from the first part of the study, categorizing each interaction involved in the creation of infographics to the six design tasks of interest in Section~\ref{sec:tasks}. Interactions that were not part of the six categories were classified as 'others' category. In the coding process, the two authors sat together and divided the video into segments. Each segment refers to a sequence of interactions related to the same category. Note that in the PowerPoint group, there are several design categories, such as `Pivot Figure Design', and `Visual Element Design', that may cover a sequence of interactions ranging from searching the resource online, uploading to PowerPoint, and designing. For each segment, its \textit{start time}, \textit{end time} and \textit{category} were identified and saved. 
A total of 32 interaction logs (2 systems * 8 participants * 2 infographics) were analyzed. We then adopted the timeline-based visual analytics method~\cite{guo2015case} to examine the time cost and temporal patterns in the interaction logs in both the GraphiMind group and the PowerPoint group.

\begin{figure*}[tbp]
    \centering    \includegraphics[width=0.9\textwidth]{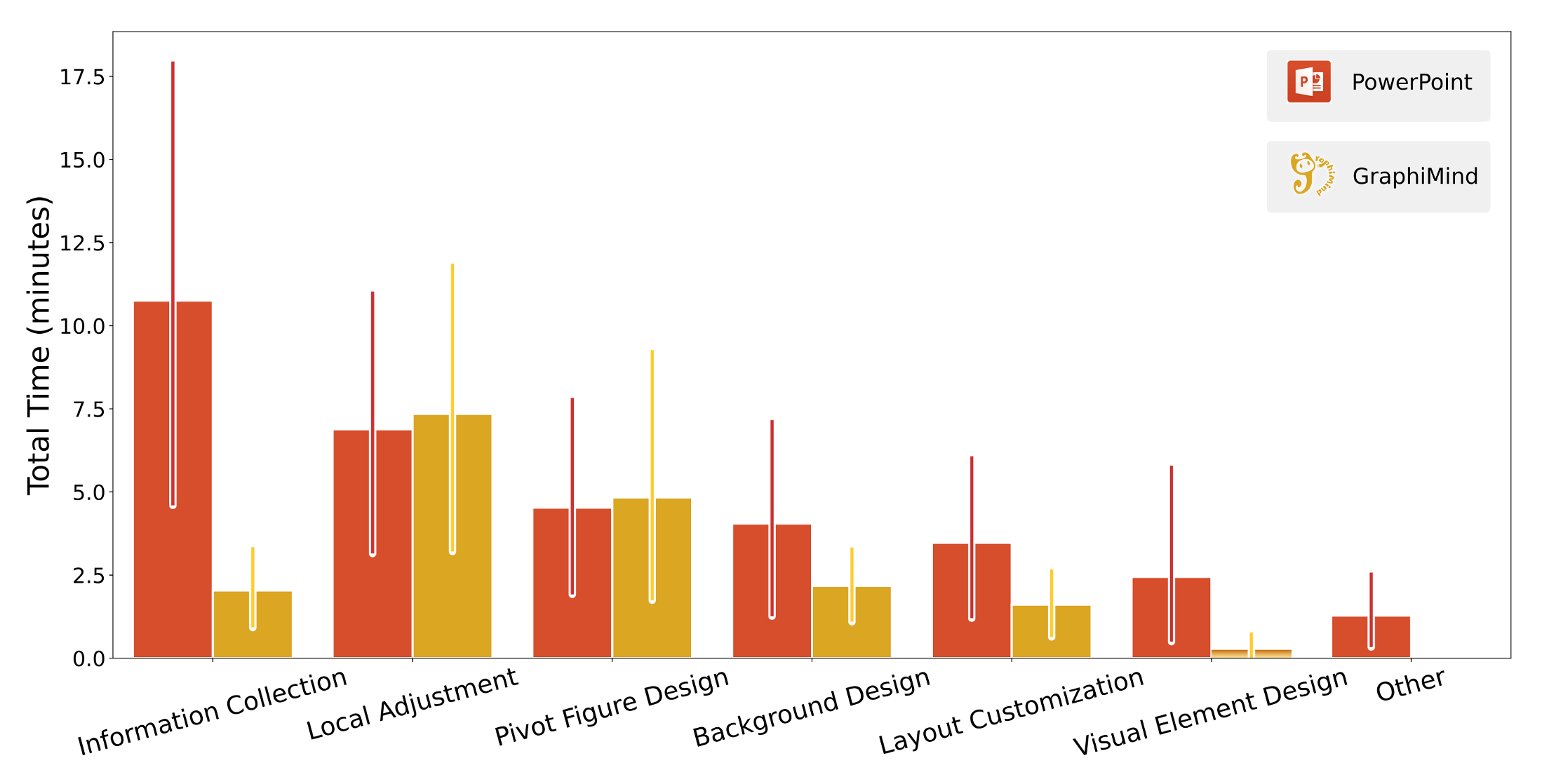}
    \caption{\cmark{Average time spent on 
    interaction categories using PowerPoint (red) and GraphiMind (yellow). The bars represent the mean time duration per category, with error bars indicating standard deviation among users.}}
    \label{Fig - Average Time per Operation}
\end{figure*}

\begin{figure*}[htbp]
    \centering    \includegraphics[width=\textwidth]{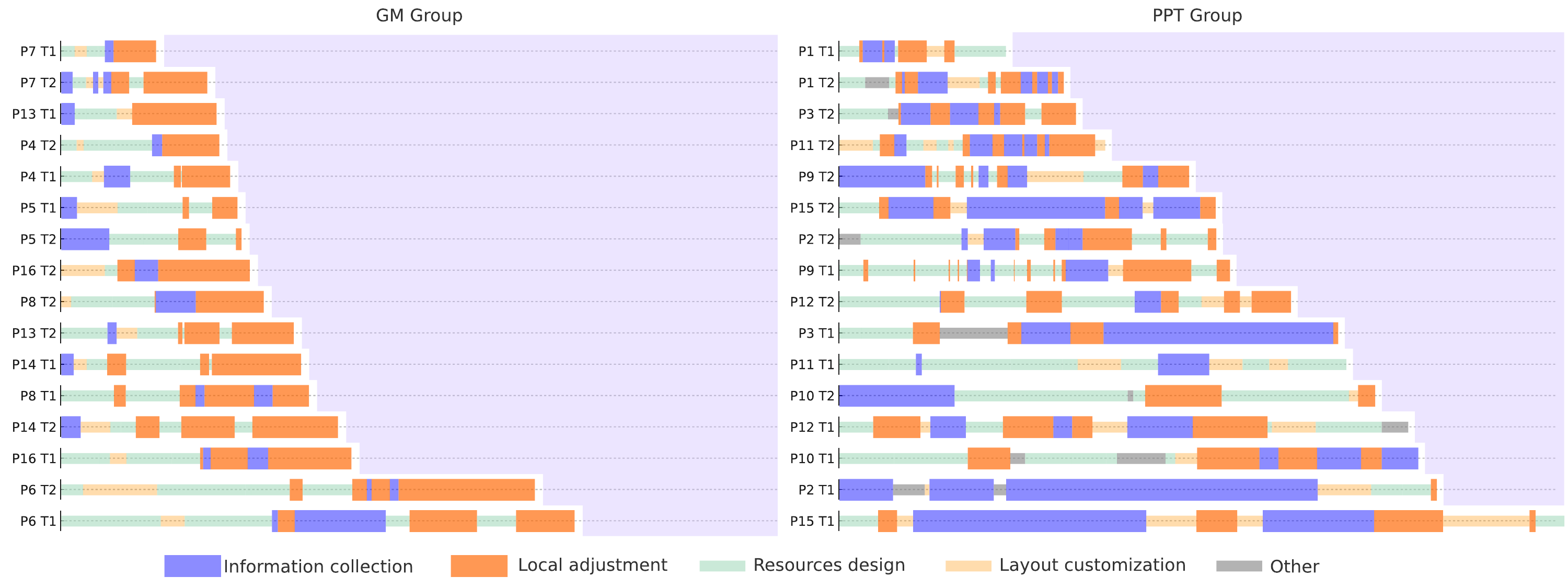}
    \caption{\cmark{Interaction Sequences of the GraphiMind and PowerPoint Groups: Each row depicts the interaction sequences of a participant in designing infographics for one of the topics. Segments in a sequence row represent the time cost of corresponding interactions. In contrast to the PowerPoint group (on the right), GraphiMind (on the left) demonstrates a more streamlined workflow.}}
    \label{Fig - Operation Sequence per Example}
\end{figure*}

\subsection{Results}

We first present the interact analysis, comparing how participants used GraphiMind to the use of PowerPoint when designing information graphics. Then, we present the questionnaire results and the analysis of the open-ended questions to provide more details on participants' opinions of its use.
\cmark{The example outcomes are shown in Figure~\ref{Fig - Visual Quality}.}

\subsubsection{Interaction Analysis} \paragraph{Interaction Time} 
Figure~\ref{Fig - Average Time per Operation} shows the average time spent on the different operations for each of the two systems. Overall, it took less time for participants to complete a satisfactory graphical design with GraphiMind ($M = 18.26$ minutes, $SD = 8.86$) than with PowerPoint ($M = 33.40$ minutes, $SD = 12.24$). This difference was significant, $t(14) = 3.28$, $p < 0.01$. Among the six interaction categories, GraphiMind saved the most time on information collection. On average participants took 2.03 minutes to collect topic-related information using GraphiMind compared to 10.76 minutes ($SD = 9.47$) with PowerPoint, which was also the most time-consuming task among the six for the PowerPoint group. This difference was significant, $t(14) = 3.37$, $p < 0.01$.
Reduction in time cost was also observed in the design tasks of background design, layout customization, and visual element design, although these differences were not significant.

\paragraph{Patterns in Design Workflow} Figure~\ref{Fig - Operation Sequence per Example} visualizes the sequences of 32 interaction logs according to the two systems used. For simplicity, the interaction categories `Pivot Figure Design', `Background Design', and `Visual Element Design' are merged into a meta-category `Resource Design'. When looking at the interaction sequences in Figure~\ref{Fig - Operation Sequence per Example}, several insights can be seen. First, the overall shortened time when using GraphiMind is evident. Apart from participant P6, GraphiMind facilitated the shorter design process for the participants. Second, the workflow of the GraphiMind group appears to be more streamlined and sequential than the PowerPoint group, i.e., starting from resources or layout design, followed by information collection, ending with local adjustment. In contrast, the PowerPoint group exhibits a more fragmented process, interspersed with frequent local adjustments (i.e., the orange segments in Figure~\ref{Fig - Operation Sequence per Example}) and information collection (i.e., the purple segments). Finally, it can be seen that a long use of local adjustments occurred at the end of the session for almost all the participants in the GraphiMind group, while in PowerPoint local adjustments were scattered throughout the whole process. 
This can be explained by the need to perform adjustments after design resources were created or imported into PowerPoint. In contrast, in GraphiMind, these resources were seamlessly integrated and fitted into a selected layout on the canvas, potentially reducing time and workload.

}

\begin{figure*}[tbp]
    \centering    \includegraphics[width=0.9\textwidth]{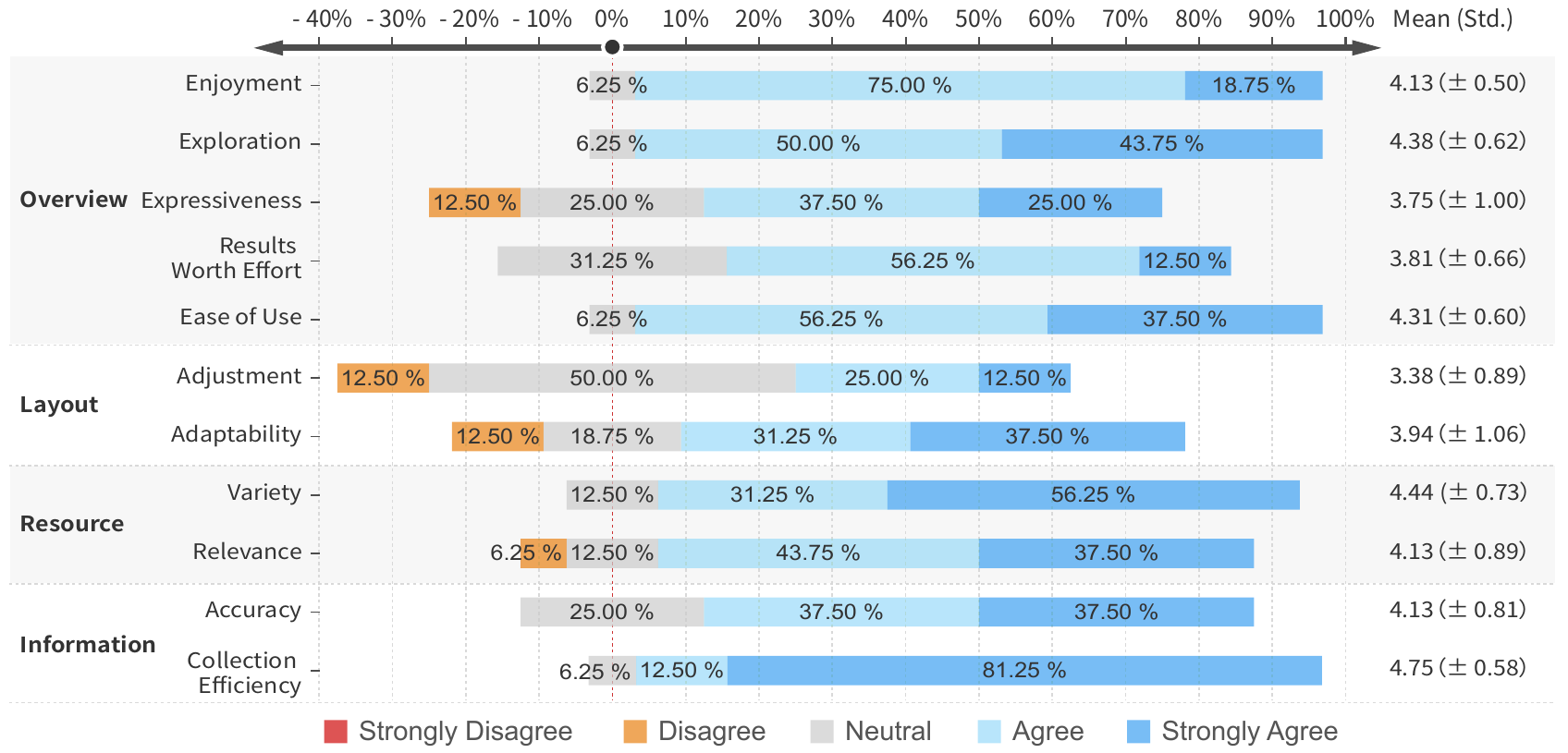}
    \caption{\cmark{The User Feedback for the 11 Items. The chart visualizes the responses from participants based on a 5-point Likert scale. Each horizontal bar represents the percentage of participants' level of agreement from 'Strongly Disagree' to 'Strongly Agree'.}}
    \label{Fig - 5 likert chart}
\end{figure*}

\subsubsection{Questionnaire}

\cmark{
The analysis of the 11 Likert-scale questions reveals overall high scores across most items, as presented in Figure \ref{Fig - 5 likert chart}, indicating a positive overall reception. Notably, 'Information Collection Efficiency' received the highest rating ($M = 4.75$ ± 0.58), aligning with findings from the first part where GraphiMind demonstrated a significant advantage in information collection. Conversely, scores related to layout items, i.e., 'Layout Adjustment' ($M$ = 3.38 ± 0.89) and 'Layout Adaptability' ($M$ = 3.94 ± 1.06), were relatively lower. This is also consistent with observations from the first part, where GraphiMind users spent more time on local adjustments compared to the PowerPoint group. 
}

\subsubsection{Open-ended Questions}

\cmark{

Overall, participants thought that GraphiMind was beneficial in making the information graphic design process simple and helpful. Here are the main themes that were extracted from the open-ended question: "What features in the system do you think help create infographics?" The focus here is on identifying and analyzing the specific features that users found most beneficial and why they are considered helpful in the infographic creation process.

\textbf{Smooth Integration between Text and Canvas.}
Participant's responses highlighted that GraphiMind's integration of a textual conversational interface and a graphical manipulation interface significantly streamlines and simplifies the infographic creation process. P2 highlighted the natural interaction, stating, \textit{"The interactive method is very convenient, switching between left and right views feels natural."} P6 praised the ease of layout design and drag-and-drop functionality, emphasizing, \textit{"The direct drag-and-drop functionality and the overall fit-in from left to right views enable quick infographic creation."} P11 discussed the user-friendly manipulation of elements, saying, \textit{"I can easily drag and drop text and images from the chat box, it is cool!"} demonstrating the software's ability to facilitate a smooth and intuitive design workflow.

\textbf{Effortless Launch of Design Process.} GraphiMind offers a clear and easy start for the infographic design process. Users can easily experiment or get inspired with new ideas by simply typing in textual requests in GraphiMind. P13 expressed the ease of initiating design ideas with GraphiMind, \textit{"I could just type in texts and see what is returned, by which the decision is much easier to make"}. P15 from the PowerPoint group, who also spent the longest time among all participants, discussed the difficulties faced with traditional methods, noting, \textit{"It [GraphiMind] greatly improves the efficiency of searching for information [...] It is quite hard to settle down on a concrete idea when I surfed the Internet"}. 

\textbf{Efficient Design Resource Generation.} GraphiMind excels in its precise generation of key visual design resources such as images, icons, and layouts. For example, P5 highlighted the system's ability to automatically create layouts that effectively meet design requirements, \textit{"Auto-generating layout and corresponding information is quite helpful."} P16 emphasized the speed at which GraphiMind generates visual elements, \textit{"It [GraphiMind] generates images, icons quickly, providing relevant visual resources."} Additionally, participants appreciated the accuracy of GraphiMind's design resource suggestions. This is particularly evident in the context of thematic image generation, as P14 mentioned, \textit{"The system can provide accurate suggestions on thematic images, which are very much in line with my expectations."} 

\textbf{Easy Textual Information Collection.} Participants reported that GraphiMind significantly simplifies the collection of textual information for infographics. Compared to conventional information searching such as by Google or other websites, the information collection functionality powered by ChatGPT in GraphiMind can collect the information related to a topic from all sources in one step, which was highly valued by participants. For example, P3 highlighted the efficiency of content generation tools, \textit{"The bulk content generation tools help me quickly obtain reasonable content."} Additionally, P8 praised the tool for its effectiveness in summarizing text information, \textit{"The completeness of the text information summary is high, the content is relevant, saving a lot of time in researching materials"}. 

\textbf{Natural Conversational Interaction.} It stands out for its intuitive and natural communication medium, leveraging language, which is inherently a primary mode of human interaction. This attribute significantly enhances its user-friendliness, especially for beginners. Users can articulate their design needs in simple, conversational language, which aligns with the natural human propensity for verbal communication, thereby reducing the learning curve and cognitive load in stark contrast to traditional graphic design interfaces. \cmark{Participant feedback highlights its advantage. P1 noted its ease for beginners: "It [dialogue interaction] is user-friendly for beginners, no need to understand specific tools, more relaxed, content generation," highlighting its accessibility. P16 found it creatively stimulating: "Dialogue interaction provides more inspiration,"}

}

\cmark{
\subsection{Discussion} 

Results of the study indicate a clear time benefit for GraphiMind over the baseline PowerPoint system. Participants completed their designs faster with GraphiMind, with a clear benefit in the information collection task. In only two categories GraphiMind did not take less time than PowerPoint:  `pivot figure design' and `local adjustment'. For pivot figure design, looking at the videos, we observed that participants could easily find pivot figures  with Google Images or other image bank websites when it came to a real-world topic. For example, it was easy to find a photo of Michael Jackson as a pivot figure in the topic of Favorite Character. However, as GraphiMind uses generative images, it is difficult for it to generate precise photo-realistic images of specific individuals. Text-to-image deep learning models can more effectively handle pivot figures on topics for which real-world photos may be unavailable. Examining the local adjustment category, The main reason it took slightly more time with GraphiMind is that GraphiMind is still a prototype system, and its canvas functions are less professional. For example, it lacks alignment and undo functionalities. More advanced canvas features can easily be integrated into GraphiMind in the future.

Overall, participants appreciated the use of Graphimind indicating high levels of enjoyment, exploration, and ease of use. In addition, participants indicated that the system provided relevant and diverse text and design resources, easing the start of the information graphic design process, and allowing for precise and easy generation of visual design resources. There are also some insights on which GraphiMind can be improved in the future: 

\paragraph{Awareness of Global Context and Design Preference} In the current implementation of GraphiMind, although it keeps track of conversions between users in the chatbox, the design resources and interactions in the canvas view have not been well-tracked yet. Several participants in the study indicated the need for \textbf{design personalization}, that is, providing services that better align with user expectations and habits. For example, P3 emphasized the need for the system to understand individual design preferences, saying, \textit{"I hope the system can align to my personalized design preferences especially when it already has a record of my historical usage"}. Other participants expressed their desire for GraphiMind to exhibit a more comprehensive perception of global contextual information (i.e., \textbf{global context awareness}). For instance, P12 specifically suggested the system to \textit{recommend theme images based on the color scheme of the background image}. This implies that the generation of theme images should consider not only the input caption but also other existing design resources to maintain consistency in the design. Furthermore, another aspect of global context awareness involves GraphiMind's sensitivity to the canvas state. For example, if a user selects a visual element on the canvas and inputs text like “It should be...”, an ideal response from GraphiMind would be to recognize that "it" refers to the selected element, thereby facilitating a smoother design experience. 

\paragraph{Describe Precision in Textual Design} One key idea of this work is to use textual conversations for information graphic design, via which users can design through words. However, there is also a shortcoming of using textual conversation for graphical design, i.e., its lack of precision and control over intricate design details. While language is a tool readily accessible to most people, the challenge arises in using it to precisely express different ideas. Furthermore, the inherent ambiguities and potential for misinterpretation in language can lead to challenges in conveying complex design ideas. Users often encounter limitations in making fine-tuned adjustments or in accurately translating complex design concepts into textual descriptions. How to improve the precision of textual description in design resources will be an interesting research topic to study.

\paragraph{Richer Design Resource Recommendations} GraphiMind demonstrates a core set of design resource recommendations, including pivot figures, visual elements, background, and layout. More flexible design resources shall be considered in the future, which can further ease the user's design tasks. For example, participant P16 pointed out his need, stating, \textit{"I hope it can recommend more things, such as font and color palette."}. Some other participants asked for recommendations on visual effects or transferred styles. GraphiMind serves as a basic LLM-centric framework from which it can be easily extended for more powerful design resource recommendations. Large language model improvements in the quality or efficiency of these models will enhance GraphiMind's overall performance.

Finally, given that GraphiMind is a prototype system, it naturally presents several opportunities for incremental engineering enhancements that can improve the usability and usefulness of GraphiMind. These may include advanced canvas features like alignment and undo functionality for refined design control. The implementation of a file system to save and retrieve work progress was also suggested for better workflow management. Participants also asked for enhanced artistic expression through text effects and multi-layer features, augmenting design aesthetics. Additionally, a sidebar thumbnail overview for quick access to the conversation history was proposed to facilitate easier navigation and quick reference.
Finally, it is important to note that GraphiMind's performance is intrinsically linked to its underlying deep learning models, such as the large language model driving its reasoning capabilities and the Stable Diffusion model enabling its text-to-image functionality.
Therefore, any improvements in the quality or efficiency of these models will directly enhance GraphiMind's overall performance.

}
\section{Conclusions}


\cmark{This work introduces \textit{GraphiMind}, an LLM-centric interface that leverages the power of the Large Language Model to facilitate the information graphics design for novice users. The core concept involves integrating a textual conversational interface into the design process, enabling users to create designs using natural language. Through an exploration of information graphics design tasks and the implementation of state-of-the-art AI models, GraphiMind supports essential design tasks by seamlessly bridging the gap between text and canvas in the design workflow. Our user study examined the interaction patterns and performance of this new system.
Compared to conventional design baseline tools, the results show that the LLM-centric conversational interface significantly streamlines the design process and enhances the efficiency and user experience of creating graphics, bringing benefits such as easy initiation of design ideas, efficient information collection, and more.

As a foundational LLM-centric framework for information graphic design, GraphiMind paves the way for extensive exploration in the realm of textual graphic design. A direct extension involves expanding the third-party tools to enhance the system's versatility. Another intriguing avenue for future development is augmenting the agent's capabilities to enable it to comprehend a broader interaction context, including the non-textual canvas state. This contextual awareness would empower the agent to offer more relevant design recommendations and automated solutions. Furthermore, significant progress could be achieved through user personalization, allowing the agent to adapt to individual user preferences and styles over time. This not only fosters a more personalized user experience but also streamlines future design workflows by aligning closely with the user's unique needs and aesthetics.}

{
    \small
    \bibliographystyle{ieeenat_fullname}
    \bibliography{main}
}


\end{document}